\documentstyle{article}
\begin{document}

\newcount\sectionnumber
\sectionnumber=0
\def\be{\begin{equation}}
\def\ee{\end{equation}}

\title{Higher Dimensional Operators and Low Energy Left-Right Symmetry}
\author{Alakabha Datta${}^{a)}$, Sandip Pakvasa${}^{a)}$ and
Utpal Sarkar${}^{b)}$}
\date{\mbox{}}

\maketitle
\begin{center}
${}^{a)}$ {\it Physics Department, University of Hawaii at
Manoa, 2505 Correa Road, Honolulu, HI 96822, USA.}\\

${}^{b)}$ {\it Theory Group, Physical Research Laboratory,
Ahmedabad - 380009, India.}\\

\end{center}

\begin{abstract}
 We consider higher dimensional operators due to quantum
gravity or spontaneous compactification
of extra dimensions in Kaluza-Klein type theory and their effect
in the $SO(10)$ Lagrangian. These operators change the boundary
conditions at the unification scale. As a result one can allow
left-right symmetry to survive till very low energy (as low as
$\sim$ TeV) for a wide range of values for the coupling of these
higher dimensional operators and still make the theory
compatible with the latest values of $\sin^2 \theta_W$ and
$\alpha_s$ derived from LEP. We consider both non-supersymmetric
and supersymmetric cases with standard higgses. Proton lifetime
is very large in these theories

\end{abstract}

The precision measurements \cite{lep} at LEP has put constraints
\cite{const,alt,amaldi,lrgut} on many extensions of the standard
model. Grand Unified Theories also fall under these constrained
category \cite{amaldi,lrgut}. The measurements of the $Z$ mass
and width and also the jet cross-sections and the various
asymmetries provide very accurate values for the
$\sin^2\theta_w$ and $\alpha_s$ at the scale $M_Z$. Using these
experimental values\cite{lep,amaldi},
\begin{eqnarray}
\sin^2 \theta_W &= &0.2333 \pm 0.0008 \nonumber \\
\alpha_s &= &0.113 \pm 0.005 \label{eqn1}
\end{eqnarray}
and taking the fine structure constant at the electroweak scale
to be, $\alpha_{em}(M_z) = 1/127.9$, one can write down the
values \cite{amaldi} of the three coupling constants at the electroweak scale
$(M_z)$.  Then using the evolution of these coupling constants
with energy it is possible to see if the coupling constants meet
at a point giving rise to grand unification of all the three
forces \cite{amaldi}. It was found that any GUT without
supersymmetry with no
intermediate mass scale are ruled out  by this analysis, whereas
theories with intermediate mass scale \cite{lrgut} get
constraints on the allowed values for the intermediate mass
scale.

An important class of GUTs with intermediate symmetry breaking
scale is the one containing those with left-right symmetry
\cite{lr}. In these theories the standard electroweak model
$SU(2)_L \otimes U(1)_Y$ emerge at low energy as a result of
symmetry breaking of a larger left-right symmetric $SU(2)_L
\otimes SU(2)_R \otimes U(1)_{B-L}$ group. The strongest bound
on this left-right symmetry breaking scale $M_R$ comes from an
analysis of the LEP data and that is only about a TeV
\cite{alt}. However, if this theory is embedded in a GUT, then
the constraint becomes \cite{lrgut},
\be
        M_R \geq 10^{9} GeV.
\ee
Here we address the question if
the right handed gauge bosons are seen in the near future in SSC
or LHC, then will that rule out the possibility of GUT ?

Attempts have been made to understand this problem
\cite{par,des}. It was found that if one breaks the
left-right D parity  spontaneously, then one can have
$g_L \neq g_R$. In this case the higgs sector is left-right
asymmetric and one can have low $M_R$. But for $M_R \sim$ TeV,
it is required to break the $SU(2)_R \rightarrow U(1)_R$ at a high
scale and as a result one can have a light $Z_R$ but not
light right handed charged gauge bosons \cite{par} naturally. To
get all the right handed gauge bosons light one requires large
numbers of artificial higgs scalars, which are left-right
asymmetric.  In the supersymmetric $SO(10)$ theory it is
possible to have low $M_R$ for a very specific choice of higgs
scalars \cite{des}.

In the case of minimal $SU(5)$ GUT it was argued that the addition
of nonrenormalizable terms induced by quantum gravity or by the
compactification of the higher dimensions (in some theories like
Kaluza-Klein theories) can allow a range of parameters
\cite{hidim,dimsix,kk} for which the correct
values for the $\sin^2\theta_w$ and $\alpha_s$ are reproduced
and which remain consistent
with  proton decay. We shall consider the higher dimensional
operators in a $SO(10)$ version of the GUT and see if for some
range of values for the coupling constants of the higher
dimensional operators we can have low value of the $M_R$
consistent with $\sin^2\theta_w$ , $\alpha_s$ and  proton
decay.

In the case of $SU(5)$ it was found that by inclusion of
the dimension five operators \cite{hidim} alone it is not
possible to get correct value of the $\sin^2\theta_w$ and
$\alpha_s$, but by including the dimension six operators
\cite{dimsix} along with the dimension five operators it is
possible to solve the problem. In the case of $SO(10)$ GUT we
find that the dimension six operators do not play any role and
if we have only two stages of symmetry breaking then we can not
have low energy left-right symmetry breaking, but if one breaks
$SO(10)$ in two stages near the Planck scale to the left-right
symmetric model then it is possible to have low energy
left-right symmetry breaking with a wide range of the values of
the coupling constants.

We consider the symmetry breaking chain to be,
\begin{eqnarray}
SO(10) \:\:\:\:\: {M_U \atop \longrightarrow}& SU(4) \times SU(2)_L \times
SU(2)_R & \left[ \equiv G_{PS} \right] \nonumber \\
&& \nonumber \\
{M_1 \atop \longrightarrow}& SU(3)_c \times SU(2)_L \times SU(2)_R \times
U(1)_{(B-L)} & \left[ \equiv G_{LR} \right] \nonumber  \\
&& \nonumber\\
{M_R \atop \longrightarrow}& SU(3)_c \times SU(2)_L  \times
U(1)_Y & \left[ \equiv G_{std} \right] \nonumber \\
&& \nonumber\\
{M_W \atop \longrightarrow}& SU(3)_c \times U(1)_{em}.& \nonumber
\end{eqnarray}
The $SO(10)$ invariant lagrangian, which allows the above
symmetry breaking chain, in the domain of energies near the
Planck scale is given as a combination of the usual four
dimensional terms and the new induced higher dimensional
nonrenormalizable terms.
These higher
dimensional terms will be suppressed by the Planck scale
($M_{Pl}$) (or by the compactification scale which can even be
two orders of magnitude below the Planck scale \cite{kk}). The
lagrangian can then be written as,
\be
L=-{1 \over 2} {\rm Tr} (F_{\mu\nu} F^{\mu\nu})
 +\sum_{n=1} L^{(n)} \label{eqn2}
\ee
and the sum in Eq.~\ref{eqn2} runs over all possible higher
dimensional operators. We write down the five- and
six-dimensional operators explicitly as
\be
L^{(1)}=-{1 \over 2} {\eta^{(1)} \over M_{Pl}} {\rm Tr} (F_{\mu\nu}
\phi F^{\mu\nu}) \label{eqnl1}
\ee
\begin{eqnarray}
L^{(2)}&=&-{1 \over 2} {1 \over M_{Pl}^2} \biggl\lbrack
\eta_a^{(2)} {\rm Tr} (F_{\mu\nu} \phi^2 F^{\mu\nu})
+ \eta_b^{(2)} {\rm Tr}(\phi^2){\rm Tr} (F_{\mu\nu} F^{\mu\nu})
\nonumber \\ && +
\eta_c^{(2)} {\rm Tr} (F^{\mu\nu}\phi) {\rm Tr} (F_{\mu\nu}
\phi) \biggr\rbrack \label{eqn21}
\end{eqnarray}
In the above equations $\eta^{(n)}$ specify the couplings of the
higher dimensional operators.

Let us first consider the case when only the scale $M_U$ is
large and $M_I$ is some intermediate symmetry breaking scale,
which does not receive any contribution from the higher
dimensional operators. Suppose the $SO(10)$is broken by a
54-plet of higgs field $\Sigma$. $\Sigma$ is a traceless
symmetric field of the $SO(10)$ and the {\it vev}s of $\Sigma$
which can mediate this symmetry breaking to the group $G_{PS}$,
is given by,
\be
        \langle \Sigma \rangle = \displaystyle
\frac{1}{\sqrt{30}}\:\: \Sigma_0 \:\: {\rm diag}(1, 1, 1, 1, 1,
1, -\frac{3}{2}, -\frac{3}{2}, -\frac{3}{2}, -\frac{3}{2}).
\ee
where,
$\Sigma_0=\sqrt{6 \over 5\pi\alpha_G} M_U$ and
$\alpha_G=g_0^2/4\pi$ is the GUT coupling. Note that this has
the larger symmetry $O(6) \otimes O(4)$. If we now consider
terms of order dimension five only, then the $G_{PS}$ invariant
lagrangian will be given by,
\[
-\frac{1}{2}(1+\epsilon_4) \: {\rm Tr} (F_{\mu
\nu}^{(4)}\:F^{(4)\mu \nu}) -\frac{1}{2}(1+\epsilon_{2}) \: {\rm
Tr} (F_{\mu \nu}^{(2L)}\:F^{(2L)\mu \nu})
\]
\be
-\frac{1}{2}(1+\epsilon_{2}) \: {\rm Tr} (F_{\mu
\nu}^{(2R)}\:F^{(2R)\mu \nu}  )
\ee
Now defining the physical gauge fields below the unification
scale to be $A_i^{\prime} = A_i \sqrt{ 1 + \epsilon_i}$, we
recover the usual $G_{PS}$ lagrangian with modified coupling
constants
\begin{eqnarray}
g_4^2(M_U)&=& \bar{g}_4^2(M_U) (1 + \epsilon_4)^{-1}\nonumber\\
g_{2L}^2(M_U)&=& \bar{g}_{2L}^2(M_U) (1 + \epsilon_2)^{-1}\nonumber\\
g_{2R}^2(M_U)&=& \bar{g}_{2R}^2(M_U) (1 + \epsilon_2)^{-1}
\end{eqnarray}
The couplings $\bar g_i$ are the couplings that would have
appeared in the absence of the higher dimensional operators,
whereas the $g_i$ are the physical couplings which are evolved
down to lower scales.

We introduce the parameter
$\epsilon^{(n)}$ associated with a given operator of dimension
$n+4$ in the following way:
\be
\epsilon^{(n)}= \biggl\lbrack {1 \over \sqrt{30}} {\phi_0  \over
M_{Pl}}\biggr\rbrack^n \eta^{(n)}
\ee
We then have
\be
\epsilon^{(n)}= \biggl\lbrack \biggl\lbrace{2 \over
25\pi\alpha_G} \biggr\rbrace^{1 \over 2} {M_U  \over
M_{Pl}}\biggr\rbrack^n \eta^{(n)}
\ee
The change in the coupling constants are, $\epsilon_4 =
\epsilon^{(1)}$ and $\epsilon_2 = -\frac{3}{2} \epsilon^{(1)}$.
It is to be noted that the $SU(2)_L$ and $SU(2)_R$ always
receive equal contributions. As a result the effect of the
higher dimensional terms will be to shift the relative
contributions between the $SU(2)$s and the $SU(4)$ gauge coupling
constants (overall shifts of all the coupling constants do not
contribute to the predictions of $\sin^2\theta_w$ and
$\alpha_s$). This relative shift can be parametrized by only one
parameter, which is related to $\epsilon^{(1)}$. Thus if by
considering only dimension five terms we can not allow low
$M_R$, then dimension six or other higher dimensional operators
will also not allow low $M_R$. Hence it suffices to consider only
dimension 5 terms.

As discussed above, the effect of higher dimensional terms will
change the boundary conditions and at the unification scale
$M_U$ we have,
$$
\bar{g}_4^2 = \bar{g}_{2L}^2 = \bar{g}_{2R}^2 = g_0^2
$$
or equivalently,
\be
 {g}_4^2(M_U) (1 + \epsilon_4) =  {g}_{2L}^2(M_U) (1 +
\epsilon_2) =  {g}_{2R}^2(M_U) (1 + \epsilon_2) = g_0^2. \label{eqn24}
\ee
Using the matching conditions at the scales $M_I$,
\be
g_{3c}^{-2}(M_I) = g_{1(B-L)}^{-2}(M_I) = g_4^{-2}(M_I)
\ee
and $M_R$,
\begin{eqnarray}
g_{1Y}^{-2}(M_R) & = & {3 \over 5} g_{2R}^{-2} (M_R) +
{2 \over 5} g_{1 (B-L)}^{-2} (M_R) \nonumber \\
g_{2L}^{-2}(M_R) & = &  g_{2R}^{-2} (M_R)  \label{eq2}
\end{eqnarray}
and the evolution of the coupling constants\cite{gut},
\begin{equation}
\mu \frac{d \alpha_i(\mu)}{d \mu} =  2 \beta_i \alpha_i^2(\mu)
\end{equation}
where $\alpha_i = \frac{g_i^2}{4 \pi}$, and  $\beta_i =
\displaystyle{-\frac{1}{(4 \pi)}\left( \frac{11}{3} T_g[i] - \frac{4}{3}
T_f[i] - \frac{1}{6} T_s[i] \right)}$, we can now write the
following relations between the standard model coupling
constants and the unification coupling constant :
\begin{eqnarray}
\alpha_Y^{-1}(M_Z) &=& \alpha_G^{-1}(1 + \frac{3}{5} \epsilon_2 +
\frac{2}{5} \epsilon_4) + (\frac{6}{5} b_{2R} + \frac{4}{5} b_4)
M_{UI} \nonumber \\
&&+ (\frac{6}{5} b_{2R} + \frac{4}{5} b_{1(B-L)}) M_{IR} + 2
b_{1Y} M_{Rw} \nonumber \\
\alpha_{2L}^{-1}(M_Z) &=& (1 + \epsilon_2) \alpha_G^{-1} + 2
b_{2L} (M_{UI} + M_{IR} + M_{Rw}) \nonumber \\
\alpha_{3c}^{-1}(M_Z) &=& (1 + \epsilon_4) \alpha_G^{-1} + 2 b_4
M_{UI} + 2 b_{3c} ( M_{IR} + M_{Rw}) \label{eqn14}
\end{eqnarray}
where, $M_{ij} = \ln \frac{M_i}{M_j}$.

We define two quantities $A$ and $B$ by the following relations,
\begin{eqnarray}
A &=& \alpha_Y^{-1} - \alpha_{2L}^{-1} \nonumber \\
B &=& \alpha_{2L}^{-1} + \frac{5}{3} \alpha_Y^{-1} - \frac{8}{3}
\alpha_{3c}^{-1}
\end{eqnarray}
$A$ and $B$ are related to $sin^2\theta_w,\:\: \alpha, \:\: {\rm
and} \:\: \alpha_s$ by
\begin{eqnarray}
\sin^2 \theta_w &=& \frac{3}{8} - \frac{5}{8} \alpha A \nonumber
\\
1 - \frac{8}{3} \frac{\alpha}{\alpha_s} &=& \alpha B
\end{eqnarray}
Equations \ref{eqn14} may be solved for $M_{UI}$ and $M_{IR}$
in terms of $A$ and $B$ to obtain
\begin{eqnarray}
M_{UI} &=& -\frac{X}{2} + \frac{M_{Rw}}{2} - \epsilon
\frac{\alpha_{G}^{-1}}{2 b_2} \nonumber \\
M_{IR} &=& \frac{Y}{2} - 2 M_{Rw}
\end{eqnarray}
where, $X = -\frac{5 A - B}{4 b_2}$; $y = -\frac{5 A + B}{4
b_2}$; $b_2 = -\frac{11}{6 \pi}$ and $\epsilon = \epsilon_4 -
\epsilon_2$. Using the measured values of $\sin^2\theta_w$ and
$\alpha_s$ from equation 1 one gets $M_{IR} \sim 2 \times
10^{22} M_R$ for $M_R \sim 10 M_w$.

Thus we see that when the $SO(10)$ GUT breaks to the $G_{PS}$ at
the unification scale it is impossible to have low $M_R$. Let us
now consider the symmetry breaking chain in which $SO(10)$ breaks
down to $G_{LR}$ at the unification scale $M_U$ directly, when
the component of a 45-plet of higgs $H$ which is invariant under
$G_{LR}$ acquires {\it v.e.v}. This can be written as,
\be
\langle H \rangle = \displaystyle \frac{1}{\sqrt{12}\:
i} \:\: H_0 \: \pmatrix{ 0_{33} & 1_{33} & 0_{34} \cr -1_{33} & 0_{33} &
0_{34} \cr 0_{43} & 0_{43} & 0_{44} }
\ee
where, $0_{mn}$ is a $m \times n$ null matrix and $1_{mm}$ is a
$m \times m$ unit matrix. The antisymmetry of the matrix will
imply that to dimension five operators there is no contribution
from this higgs. The lowest order contribution comes from the
dimension six operators. To order six the $G_{LR}$ invariant
lagrangian is written as,
\[
-\frac{1}{2}(1+\epsilon_3) \: {\rm Tr} (F_{\mu
\nu}^{(3)}\:F^{(3)\mu \nu}) -\frac{1}{2}(1+\epsilon_{2}) \: {\rm
Tr} (F_{\mu \nu}^{(2L)}\:F^{(2L)\mu \nu})
\]
\be
-\frac{1}{2}(1+\epsilon_{2}) \: {\rm Tr} (F_{\mu
\nu}^{(2R)}\:F^{(2R)\mu \nu})  -\frac{1}{2}(1+\epsilon_{1}) \:
{\rm Tr} (F_{\mu \nu}^{(1)}\:F^{(1)\mu \nu})
\ee
where, $\epsilon_3 = \epsilon_1$ and $\epsilon_2 = 0$. As argued
earlier, in this case also there is no relative shift in the
boundary condition between the $SU(3)$ and the $U(1)$ coupling
constants and as a result it will not be possible to have low
$M_R$.

We shall now discuss the situation when the $SO(10)$ group is
broken to the $G_{LR}$ subgroup in two stages, but we now assume
that the higher dimensional operators contribute to both the
scales $M_U$ and $M_I$. Consider the five dimensional operator,
$${L^\prime}^{(1)}=-{1 \over 2} {\eta^{\prime(1)} \over M_{Pl}} {\rm Tr}
[(F_{\mu\nu} \phi_{45} F^{\mu\nu})]$$. This operator contains the
$SU(4)$ invariant piece $$ {L^{\prime\prime}}^{(1)} =
-\frac{1}{2} {\eta^{\prime(1)} \over M_{Pl}} {\rm Tr}
[(F_{\mu\nu}^{(4)} \phi_{15}^{\prime} F^{(4) \mu\nu})]$$ where,
$\phi_{15}^{\prime}$ transforms as $(15,1,1)$ under $SU(4)_c
\otimes SU(2)_L \otimes SU(2)_R$. At the scale $M_I$ one can now
break $SU(4)_c \longrightarrow SU(3)_c \otimes U(1)_{(B-L)}$ by
giving a {\it vev} to $\phi_{15}^\prime$ of the form,
\be
\phi_{15}^\prime = \frac{1}{\sqrt{24}} \phi_0^\prime {\rm diag}
[1, 1, 1, -3].
\ee
The $SU(3)_c \otimes U(1)_{(B-L)}$ invariant kinetic energy
term for the gauge bosons will now be
\be
-\frac{1}{2}(1+\epsilon_3^\prime) \: {\rm Tr} (F_{\mu
\nu}^{(3)}\:F^{(3)\mu \nu}) -\frac{1}{2}(1+\epsilon_{1}^\prime)
\: {\rm Tr} (F_{\mu \nu}^{(1)}\:F^{(1)\mu \nu})
\ee
where,
\[
\epsilon_3^\prime = [\frac{1}{\sqrt{24}}
\frac{\phi_0^\prime}{M_{Pl}} ] \eta^{\prime (1)} = \left[ \left\{
\frac{2}{25 \pi \alpha_4}\right\}^{\frac{1}{2}}
\frac{M_I}{M_{Pl}} \right] \eta^{\prime (1)}
\]
and $\epsilon_1^\prime = - 2 \epsilon^\prime_3$ (for dimension 5
operators only).

The one loop matching conditions at the scale $M_U$ is same as
in equation \ref{eqn24}, while that at the scale $M_I$ is
modified to,
\be
g_{1(B-L)}^2(M_I) (1 + \epsilon_1^\prime) = g_{3c}^2(M_I) (1 +
\epsilon_3^\prime) = g_4^2(M_I).
\ee
The other matching conditions remain unchanged.

The equations \ref{eqn14} now modify to
\begin{eqnarray}
\alpha_Y^{-1}(M_Z) &=& \alpha_G^{-1}(1 + \frac{3}{5} \epsilon_2 +
\frac{2}{5} \epsilon_4 + \frac{2}{5} \epsilon_1^\prime (1 +
\epsilon_4))  \nonumber \\
&&+ (\frac{6}{5} b_{2R} + \frac{4}{5} (1 + \epsilon^\prime_1) b_4)
M_{UI}+ (\frac{6}{5} b_{2R} + \frac{4}{5} b_{1(B-L)}) M_{IR} + 2
b_{1Y} M_{Rw} \nonumber \\
\alpha_{2L}^{-1}(M_Z) &=& (1 + \epsilon_2) \alpha_G^{-1} + 2
b_{2L} (M_{UI} + M_{IR} + M_{Rw}) \nonumber \\
\alpha_{3c}^{-1}(M_Z) &=& (1 + \epsilon_4)(1 +
\epsilon_3^\prime) \alpha_G^{-1} + 2 b_4 (1 + \epsilon_3^\prime)
M_{UI} + 2 b_{3c} ( M_{IR} + M_{Rw}) \label{eqn24}
\end{eqnarray}
One can now solve for $M_{UI}$ and $M_{IR}$
to obtain
\begin{eqnarray}
M_{UI} &=& -\frac{X}{2} + \frac{M_{Rw}}{2} - \epsilon
\frac{\alpha_{G}^{-1}}{2 b_2} \nonumber \\
M_{IR} &=& \frac{Y}{2} + y X - 2 M_{Rw} (1 + \frac{y}{2}) -
\frac{\epsilon \alpha_{G}^{-1}}{b_2} \sigma
\end{eqnarray}
where, $\epsilon = \epsilon_4 - \epsilon_2 = \frac{5}{2}
\epsilon^{(1)}; \:\: x = \frac{\epsilon_3^\prime (1 +
\frac{2}{5} \epsilon)}{\epsilon}; \:\: y = \epsilon^\prime_3
\frac{b_4 f}{b_2}; \:\: \sigma = x - y$ and $b_4f$ and $b_2$ are the beta
functions with and without fermionic contributions for the appropriate gauge
groups.
With $\frac{M_R}{M_w} = 10$ we give  in table 1 the allowed
ranges of $\epsilon_3^\prime$ and $\epsilon$ for some choices of
$M_U$ and $M_I$.Note that given $\epsilon_3\prime$ and $M_{UI}$ one can
use Eq.25 and any one of the relations in Eq.24 to solve for $\epsilon$ and
$\alpha_{G}^{-1}$.The parameters $\eta^{(1)}$ and $\eta^{\prime (1)}$
appearing in the Lagrangian can be obtained from Eq.10 and Eq.22 respectively.
\begin{table}
\caption{Allowed ranges of $\epsilon$ and $\epsilon^\prime$ for
various $M_U$ and $M_I$.}
\begin{center}
\begin{tabular}{||c|c|c|c|c||}
\hline \hline
$\epsilon_3^\prime$& $\epsilon$ &$M_I$& $M_U$& $\alpha_U^{-1}$\\
\hline
-.119 -- -.141 & .16 -- .163& $10^{18}$& $10^{18}$& 55.253
-- 54.211\\
\hline
-.153 -- -.176& .204 -- .208 &$ 10^{17}$ &$ 10^{18}$ &
56.493 -- 55.462\\ \hline
-.177 -- -.201 & .244 -- .248 &$ 10^{16}$ &$ 10^{18}$ &
58.252 -- 57.218\\ \hline
-.153 -- -..176 & .165 -- .168 & $10^{17}$&$10^{17}$& 53.658 -- 52.628\\ \hline
-.177 -- -.201 & .208 -- .212 & $10^{16}$&$10^{17}$& 55.418 -- 54.383\\
\hline \hline
\end{tabular}
\end{center}
\end{table}

The analysis can be extended in a straightforward way to
supersymmetric $SO(10)$ by making appropriate changes in the
$\beta$-functions, $\beta_i =
\displaystyle{-\frac{1}{(4 \pi)}\left( 3 T_g[i] -
T_f[i] -  T_s[i] \right)}$.For this case the form of Eq.24 and Eq.25 remain
unchanged though the values for the beta functions change.
The results for the supersymmetric $SO(10)$
neglecting the higgs contributions to the $b_i$s are shown in
table 2. The effect of higgs will change the result in the sense
that the allowed ranges for the $\epsilon$s will get slightly
modified. Similarly the higher loop contributions are also
likely to change the allowed ranges by small amount\cite{tl}. However,
none of these smaller effects can, in practice, change the
conclusion.

\begin{table}
\caption{Allowed ranges of $\epsilon$ and $\epsilon^\prime$ for
various $M_U$ and $M_I$ in the supersymmetric version.}
\begin{center}
\begin{tabular}{||c|c|c|c|c||}
\hline\hline
$\epsilon_3^\prime$&$\epsilon$&$M_I$&$M_U$&$\alpha_U^{-1}$\\ \hline
-.178 -- -.210 & .246 --.251 & $10^{18}$ & $10^{18}$ & 29.383
-- 28.771\\ \hline
-.229 -- -.264 & .317 -- .324 & $10^{17}$ & $10^{18}$ & 29.741 --
29.123\\ \hline
-.265 -- -.301 & .382 -- .389 & $10^{16}$ & $10^{18}$ &  30.425 --
29.283\\ \hline
-.229 -- -.264 & .254 -- .260 & $10^{17}$&$10^{17}$& 28.421 -- 27.804\\ \hline
-.265 -- -.301 & .324 -- .331 &$10^{16}$&$10^{17}$& 29.105 -- 28.503\\
\hline\hline
\end{tabular}
\end{center}
\end{table}

In summary, we have pointed out that the low energy restoration
of parity is not completely ruled out by the LEP data. In
principle the effect of quantum gravity can allow low energy
left-right symmetry in Grand Unified Theories with intermediate
symmetry breaking mass scales.

\subsection{Acknowledgement}
One of us (US) would like to acknowedge the hospitality of the
High Energy Physics Group of the University of Hawaii where this work
was done. Discussion with Prof A. Sanda is highly appreciated,
during which this problem was initiated.This work was supported in part
 by US D.O.E under contract DE-AMO3-76SF-00325.
\subsection{Note Added :}
For an earlier discussion on the possibility of low $M_R$ with gravitational
corrections see T.Rizzo,Phys.Lett.{\bf B142}, 163 (1984).


\begin{thebibliography}{References}
\bibitem{tl} Two loop calculations will be reported elsewhere.
\end{thebibliography}
\end{document}